\documentstyle[seceq,epsf]{ptptex}
\newcommand{\beq}{\begin{equation}}
\newcommand{\eeq}{\end{equation}}
\title{3D BEC Bright Solitons under Transverse Confinement: \\
Analytical Results with the Nonpolynomial Schr\"odinger Equation}

\author{Luca {\sc Salasnich} 
\footnote{E-mail address: salasnich@mi.infm.it}} 

\inst{Istituto Nazionale per la Fisica della Materia, Unit\`a di Milano, \\
Dipartimento di Fisica, Universit\`a di Milano, \\
Via Celoria 16, 20133 Milano, Italy} 

\abst{The Bose-Einstein condensate (BEC) of a dilute gas of bosons 
is well described by the three-dimensional 
Gross-Pitaevskii equation (3D GPE), that is a nonlinear 
Schr\"odinger equation. By imposing a transverse 
confinement the BEC can travel only in the cylindrical axial direction. 
We show that in this case the BEC with attractive interaction admits 
a 3D bright soliton solution which generalizes the text-book one, 
that is valid in the one-dimensional limit (1D GPE). 
Contrary to the 1D case, the 3D bright soliton exists 
only below a critical number of Bosons that depends on the extent 
of confinement. Finally, we find that the 3D bright soliton collapses 
if its density excedes a critical value. Our results are obtained 
by using a nonpolynomial Schr\"odinger equation (NPSE), an effective 
one-dimensional equation derived from the 3D GPE.} 

\begin{document}

\maketitle

\section{Introduction}

In a dilute gas of $N$ Bosons at zero temperature practically 
all particles are in the same single-particle state of the 
density matrix $\rho({\bf r},{\bf r}';t)$.\cite{rf:1,rf:2} 
This macroscopically occupied 
single-particle state is called Bose-Einstein condensate (BEC) 
and it is well described by a complex 
classical field $\psi({\bf r},t)$ (order parameter or 
macroscopic wavefunction) whose Lagrangian 
density is given by 
\beq
{\cal L} = \psi^* \left[ i\hbar {\partial \over \partial t} 
+{\hbar^2 \over 2 m} \nabla^2 - U \right] \psi -{1\over 2} g |\psi|^4\; ,  
\eeq 
where $U({\bf r})$ is the external potential, $g=4\pi\hbar^2a_s/m$ 
is the interatomic strength with $a_s$ the s-wave 
scattering length, and the complex field is normalized to $N$. 
By imposing the least action principle one obtains the following 
Euler-Lagrange equation 
\beq 
i\hbar {\partial \over \partial t} \psi 
= \left[ -{\hbar^2 \over 2 m} \nabla^2 + U 
+ g |\psi|^2 \right] \psi \; ,  
\eeq 
which is a nonlinear Schr\"odinger equation (NLSE) and it is called 
three-dimensional Gross-Pitaevskii equation (3D GPE).\cite{rf:1,rf:2} 
\par 
In recent experiments \cite{rf:3,rf:4} Bose-Einstein condensates 
have been trapped in quasi-1D cylindrical optical traps to study 
bright soliton solutions,\cite{rf:5} which are shape-invariant 
matter waves with attractive interatomic interaction ($a_s<0$). 
In this paper we derive analytical results for the 
BEC bright solitons by using a one-dimensional 
nonpolynomial Schr\"odinger equation (NPSE) we have recently 
obtained from the 3D GPE.\cite{rf:6,rf:7} 

\section{From 3D GPE to NPSE}

The cylindrical optical confinement of Ref. 3 and Ref. 4 
can be described by a harmonic potential 
\beq 
U({\bf r})={1\over 2}m\omega_{\bot}^2(x^2+y^2) \; , 
\eeq 
in the transverse direction of the BEC. 
This external potential suggests to map the 3D GPE into an effective 
1D equation, which simplifies greatly the solution of the 3D GPE. 
Our approach is to choose for the complex field the 
following variational ansatz 
\beq 
\psi({\bf r},t) = N^{1/2} \phi(x,y,t;\sigma(z,t)) \; f(z,t) \; , 
\eeq 
where both $\phi$ and $f$ are normalized to one and $\phi$ is represented 
by a Gaussian:  
\beq
\phi(x,y,t;\sigma(z,t)) = { e^{-(x^2+y^2)\over 2 \sigma(z,t)^2} 
\over \pi^{1/2} \sigma(z,t)} \; . 
\eeq
Moreover we assume that the transverse wavefunction $\phi$ is slowly 
varying along the axial direction with respect 
to the transverse direction, 
i.e. $\nabla^2 \phi \simeq \nabla_{\bot}^2 \phi$  
where $\nabla_{\bot}^2={\partial^2/\partial x^2}+  
{\partial^2/\partial y^2}$.  
By using the trial wave-function in the Lagrangian density  
and after spatial integration along $x$ and $y$ variables  
the Lagrangian density becomes a function of $\sigma$, 
$f$ and $f^*$. The Euler-Lagrange equation with respect  
$\sigma$ is given by 
\beq 
\sigma^2 = a_{\bot}^2 \sqrt{1 + 2 a_s N |f|^2} 
\; , 
\eeq 
where $a_{\bot}=\sqrt{\hbar/(m\omega_{\bot})}$ 
is the oscillator length in the transverse direction. 
Inserting this result in the Euler-Lagrange 
equation with respect to $f^*$ one finally obtains 
$$
i\hbar {\partial \over \partial t}f=  
\left[ -{\hbar^2\over 2m} {\partial^2\over \partial z^2} 
+ {g N \over 2\pi a_{\bot}^2} {|f|^2\over 
\sqrt{1+ 2a_sN|f|^2} }  \right. 
$$
\beq
\left. 
+ {\hbar \omega_{\bot}\over 2}   
\left( {1\over \sqrt{1+ 2 a_sN|f|^2} } 
+ \sqrt{1+ 2a_sN|f|^2} \right) \right] f  \; . 
\eeq  
This equation is a time-dependent non-polynomial 
Schrodinger equation (NPSE). Note that 
under the condition $a_sN|f|^2 \ll 1$ 
one has $\sigma^2=a_{\bot}^2$ and NPSE reduces to 
\beq 
i\hbar {\partial \over \partial t}f= 
\left[ -{\hbar^2\over 2m} {\partial^2\over \partial z^2} 
+ {g N \over 2\pi a_{\bot}^2} |f|^2 \right] f  \; ,  
\eeq 
where the additive constant $\hbar \omega_{\bot}$ 
has been omitted because it does not affect the dynamics. 
This equation, called one-dimensional Gross-Pitaevskii equation (1D GPE), 
is the familiar nonlinear cubic Schr\"odinger equation. 
The nonlinear coefficient $g'$ 
of this 1D GPE can be thus obtained from the nonlinear coefficient 
$g$ of the 3D GPE by setting $g'=g/(2\pi a_{\bot}^2)$. 
Note that the limit $a_s N|f|^2 \ll 1$ is precisely the condition 
for the one-dimensional regime where the healing length 
$\xi = (8\pi a_s \rho)^{-1/2}$ of the BEC is larger 
than $a_{\bot}$. In Ref. 6 and Ref. 7 we have verified 
that NPSE is very accurate in the description 
of cigar-shaped Bose condensates both in the 3D regime and 
in the 1D regime. NPSE has been recently applied in the study 
of an atom laser.\cite{rf:8} 

\section{3D BEC Bright Solitons}

Under transverse confinement and negative scattering 
length ($a_s<0$) a bright soliton\cite{rf:5} sets up when the negative 
inter-atomic energy of the BEC compensates the positive kinetic energy 
such that the BEC is self-trapped in the axial direction. 
The shape of this 3D condensate bright soliton 
can be deduced from NPSE. 
Scaling $z$ in units of $a_{\bot}$ and $t$ in units 
of $\omega_{\bot}^{-1}$, with the standard position 
\beq 
f(z,t)=\Phi(z-vt) e^{iv(z-vt)} e^{i(v^2/2 - \mu)t} \; , 
\eeq 
from NPSE one obtains the Newtonian second-order differential equation 
\beq 
\left[ {d^2\over d\zeta^2} - 2 \gamma 
{\Phi^2\over \sqrt{1-2\gamma\Phi^2} } 
+ {1\over 2} 
\left( {1\over \sqrt{1-2 \gamma \Phi^2} } 
+ \sqrt{1-2\gamma \Phi^2} \right) \right] \Phi 
= \mu \; \Phi \; ,  
\eeq 
where $\zeta =z-vt$ and $\gamma=|a_s|N/a_{\bot}$. 
A simple constant of motion of this equation is given by  
\beq 
E={1\over 2}\left({d\Phi\over d\zeta} \right)^2 
+ \mu \; \Phi^2 -\Phi^2\sqrt{1-2\gamma\Phi^2} \; ,  
\eeq 
from which one can write 
\beq 
\int {d\Phi \over 
\sqrt{2(E-\mu\; \Phi^2 + \Phi^2\sqrt{1-2\gamma\Phi^2} ) } } 
= \int d\zeta \; . 
\eeq
By imposing the boundary condition $\Phi\to 0$ for 
$\zeta \to \infty$, which implies that $E=0$, 
by quadratures one obtains the solitary bright-soliton solution 
written in implicit form 
\beq 
\zeta= {2^{-1/2} \over \sqrt{1-\mu}} \; 
arctg\left[ 
\sqrt{ \sqrt{1-2\gamma\Phi^2}-\mu \over 1-\mu } \right] 
- {2^{-1/2} \over \sqrt{1+\mu}} \; 
arcth\left[ 
\sqrt{ \sqrt{1-2\gamma\Phi^2}-\mu \over 1+\mu } 
\right] \; .   
\eeq  
Moreover, by imposing the normalization condition to 
the function $\Phi$, one has 
\beq 
(1-\mu)^{3/2} - {3\over2} (1-\mu)^{1/2} + 
{3\over 2 \sqrt{2}} \gamma = 0 \; . 
\eeq 
The normalization relates the chemical potential $\mu$ 
to the coupling constant $\gamma$, 
while the velocity $v$ of the bright soliton 
remains arbitrary. 
Note that in the 1D limit ($\gamma\Phi^2\ll 1$), 
the normalization condition gives 
$\mu= 1 - \gamma^2/2$ and the bright-soliton solution reads 
\beq
\Phi(\zeta)=\sqrt{\gamma\over 2} \; sech\left[{\gamma}\zeta \right] \; .  
\eeq 
The above solution is the text-book 1D bright soliton 
of the 1D nonlinear (cubic) Schr\"odinger equation (1D GPE). 
\par
From Eq. (3$\cdot$6) it easy to show that for $\gamma>2/3$ there are no 
solitary-wave solutions. This is a remarkable result because,  
contrary to the 3D bright soliton the widely studied 1D bright 
soliton exists (and it is stable) at any $\gamma$. 
Thus BEC bright solitons exist only below a critical number 
\beq  
N_c = {2\over 3} {a_{\bot}\over |a_s|} 
\eeq
of Bosons. Finally, we observe that Eq. (3$\cdot$5) is well defined only for 
$\gamma \Phi^2 <1/2$; at $\gamma \Phi^2 = 1/2$ the transverse 
size $\sigma$ of the BEC soliton is zero. Because $\Phi^2$ is measured 
in units of $1/a_{\bot}$, it follows that it exists a 
critical axial density 
\beq 
\rho_c = {1\over 2 |a_s|} 
\eeq  
above which there is the collapse of the BEC bright soliton. 
This result is confirmed by the numerical integration of 3D GPE. 

\section{Conclusions} 

It is well known that in one dimension soliton states exist for Bose 
condensed atoms with negative scattering length $a_s$ 
described by the one-dimensional Gross-Pitaevskii equation, that is 
the one-dimensional nonlinear cubic Schr\"odinger equation.\cite{rf:5} 
Such states, called bright solitons, are 
characterized by a time invariant shape. 
We have studied the existence
of similar states in higher dimensions. In two and three dimensions in 
a flat potential stable bright solitons do not exist.\cite{rf:5} 
However if an external 
harmonic potential constrains the motion in two dimensions 
we have found that such 3D soliton states exist 
but only below a critical number $N_c$ of bosons. The value of $N_c$ 
depends on the transverse confinement, more precisely 
it is proportional to the ratio of the transverse harmonic 
length to the scattering length. Moreover, we have found that 
3D bright solitons collapse if their axial density is larger 
than $1/(2a_s)$: for this value the transverse width of the bright 
soliton shrinks to zero. Finally, it is important to stress that 
our solitary-wave solutions can be called solitonic solutions 
because they have particle-like properties, e.g. 
collisional stability, as recently verified studying 
scattering processes.\cite{rf:9}

\end{document}